\documentstyle{laa}     
\def\ut#1{\mathop{\vtop{\ialign{##\crcr
     $\hfil\displaystyle{#1}\hfil$\crcr\noalign
     {\kern1pt\nointerlineskip}\hbox{$\hfil\sim\hfil$}\crcr
     \noalign{\kern1pt}}}}}

\def\undersymbol#1#2{\mathop{\vtop{\ialign{##\crcr
     $\hfil\displaystyle{#2}\hfil$\crcr\noalign
     {\kern1pt\nointerlineskip}\hbox{$\hfil#1\hfil$}\crcr
     \noalign{\kern1pt}}}}}
\def\arcsec{^{\prime\prime}}

\begin{document}
\thesaurus{02.07.1 ; 02.07.2 ; 12.07.1}
\title{
   Astrophysical implications of gravitational microlensing
 of gravitational waves}
\author{
    F. De Paolis,
    G. Ingrosso
\and A. A. Nucita
} \offprints{G. Ingrosso} \institute{ Dipartimento di Fisica,
Universit\`a di Lecce, and INFN, Sezione di Lecce, Via Arnesano,
CP 193, I-73100 Lecce, Italy }
\date{Received October 12, 2000; accepted November 7, 2000}
\maketitle
\begin{abstract}
Astrophysical implications of gravitational microlensing of
gravitational waves emitted by rotating neutron stars (NSs) are
investigated. In particular, attention is focused on the following
situations: i) NSs in the galactic bulge lensed by a central black
hole of $2.6\times 10^6 M_{\odot}$ or by stars and MACHOs
distributed in the galactic bulge, disk and halo between Earth
and the sources; ii) NSs in globular clusters lensed by a central
black hole of $\sim 10^3 M_{\odot}$ or by stars and MACHOs
distributed throughout the Galaxy. The detection of such kind of
microlensing events will give a unique opportunity for the unambiguous 
mapping of the central region of the Galaxy and of globular clusters. In
addition, the detection of such events will provide a new test of the 
General Theory of Relativity.
Gravitational microlensing will, moreover, increase the challenge
of detecting gravitational waves from NSs.

\keywords{Gravitation - Gravitational waves - Gravitational
lensing}

\end{abstract}

\section{Introduction}
Gravitational microlensing (i.e. gravitational amplification) of
electromagnetic waves is a well known phenomenon predicted by
the General Theory of Relativity. It is now a well-developed
observational technique in astronomy and is considered to be a
fundamental tool for acquiring information about the nature and
distribution of galactic dark matter.

Indeed, several hundreds of microlensing events have been observed
towards the galactic bulge, the Large and Small Magellanic Clouds
and even towards the M31 galaxy (see e.g. Alcock et al.
\cite{alcock}, De Paolis, Ingrosso and Jetzer \cite{dij},
Melchior, Afonso, Ansari et al. \cite{agape}). These observations
have led to the conclusion that about $50\%$ of the galactic dark
matter is in the form of MACHOs (Massive Astrophysical Compact
Halo Objects) with mass $m\simeq 0.4~M_{\odot}$. Quasar
gravitational lensing is a rather sophisticated technique and may even
allow investigators, by analyzing the rapid luminosity fluctuations in the
lensed images, to establish the nature of the dark matter in the
lens galaxies (Lewis and Belle \cite{lb}, Schmidt and Wambsganss
\cite{sw}, Mao and Schneider \cite{ms}, Gibson and Schild
\cite{gs}).

In the weak field approximation, the physics of gravitational
lensing also holds if applied not only to electromagnetic waves
(as is usually done) but  also to gravitational waves, in particular those
emitted by rotating neutron stars (NSs). This may be relevant in relation 
to the next generation of gravitational wave interferometric
detectors, such as VIRGO and LIGO (Jaranowki and Kr\`olak
\cite{jk}). In particular, these detectors will very likely be
able to detect gravitational waves emitted by rotating NSs in our
galaxy.

In the presence of a massive object (such as a star or a MACHO) close
enough to the line of sight of the observed NS, an amplification
of the received gravitational wave amplitude $h$ may be observed
under certain circumstances. Also, a massive black hole in the
center of the Galaxy or in a globular cluster center may induce
an amplification of the gravitational waves emitted by NSs situated
behind the lensing object. The aim of this paper is to investigate
some astrophysical implications of gravitational microlensing of such
gravitational waves.

We will concentrate our attention on a few astrophysical
situations which appear to be, at first sight, the most
interesting: $i$) NSs in the galactic bulge lensed by a central
black hole with mass $2.6\times 10^6 M_{\odot}$ (hereafter
indicated as BH events) or by stars and MACHOSs distributed
throughout the galactic bulge, disk and halo (hereafter indicated
as BDH events); $ii$) NSs in globular clusters lensed by a
central black hole of mass $\sim 10^3 M_{\odot}$ (again BH events)
or by stars and MACHOs distributed throughout the Galaxy (BDH
events) between Earth and the gravitational wave sources.

In analogy with the standard microlensing analysis (Schneider,
Ehlers and Falco \cite{sef}), when a lens of mass $m$ is
sufficiently close to the line of sight of a distant source, the
gravitational waves emitted by the source suffer a gravitational
deflection and the original gravitational wave luminosity
increases by the factor
\begin{equation}
A=\frac{u^2+2}{u(u^2+4)^{1/2}}~. \label{eq:bb}
\end{equation}
Here $u=d/R_E$ ($d$ is the distance of the lens from the line of
sight to the source) and $R_E$ is the Einstein radius defined as:
\begin{equation}
R_E=\left[\frac{4GmD}{c^2} x(1-x)\right]^{1/2}~,
\end{equation}
where $x=s/D$, $D$ is the distance between the observer and the
source, $s$ is the observer-lens distance. Due to the relative
transverse velocity $v_{\bot}$ between the lens and the source
(with respect to the line of sight), the amplification factor $A$
is be time-dependent and the duration time scale of the
microlensing event is given, as usual, by
\begin{equation}
\Delta T = \frac{2R_E}{v_{\bot}} \label{duration}
\end{equation}
Two other quantities are relevant for microlensing and will be
used in the following discussion: the optical depth $\tau$, i.e. the
fraction of the sky covered by the Einstein circles of all the
lenses between the observer and the sources, and the microlensing
rate $\Gamma$, which are defined, respectively, as:
\begin{equation}
\tau=\int \pi R_E^2 n(x) dx~, \label{tau}
\end{equation}
\begin{equation}
\Gamma=2\int\int R_E n(x) v_{\bot} f(v_{\bot})dx dv_{\bot}~,
\label{gamma}
\end{equation}
where $n(x)$ is the number density of the considered distribution
of lenses and $f(v_{\bot})$ is the Maxwellian transverse velocity
distribution. For lenses located in the halo or in the bulge of
the Galaxy, this function assumes the form
$\displaystyle{f(v_{\bot})=2v_{\bot}\sigma
^{-2}\exp(-v_{\bot}^2/\sigma^2)}$,  where $\sigma$ is the
one-dimensional velocity dispersion  which turns out to be $210$
km$~$s$^{-1}$ and  $156$ km$~$s$^{-1}$ for halo and bulge objects,
respectively (De R\'ujula, Jetzer and Mass\`o \cite{djm}). In the
case of lenses located in the galactic disk we have to take into
account the relative velocity $v_t$ of the observer and the
source transverse to the line of sight. The distribution of the
transverse velocity for the disk lens is thus
$\displaystyle{f(v_{\bot})=
v_{\bot}\sigma^{-2}\exp(-(v_{\bot}-v_t)^2/\sigma^2)}$, with
$\sigma\simeq 30$ km s$^{-1}$. In each considered case the
integrals on the right hand side of equations (\ref{tau}) and
(\ref{gamma}) will be performed by taking into account the real
geometry of the considered system. The last relevant quantity for
a microlensing experiment is the number of events, $N_{ev}$, detected
by observing $N$ neutron stars within an observation time $t_{\rm
obs}$, which is given by
\begin{equation}
N_{ev}= \left\{\begin{array}{ll} \Gamma \max(\Delta T,~t_{\rm
obs})~~~~~~{\rm for ~BH~ microlensing}\\
\Gamma N \max(\Delta T,~t_{\rm obs})~~~{\rm for ~BDH~
microlensing}~.  \end{array} \right. \label{nev}
\end{equation}
The matter distributed in our galaxy is assumed to follow a 
three-component model consisting of a central bulge, a double
exponential disk and a dark spherical halo. In particular, the
central concentration of stars is described by a triaxial bulge
model with a mass density law (Dwek, Arendt, Hauser et al.
\cite{dwek})
\begin{equation}
\rho_C(x,y,z) =\frac{M_b}{8\pi abc}e^{-s^2/2}~, \label{rho_c}
\end{equation}
with \begin{equation} s^4=(x^2/a^2+y^2/b^2)^2+z^4/c^4~,
\label{eq:2.1}
\end{equation}
where the bulge mass is $M_b \sim 2 \times 10^{10}~M_{\odot}$ and
the scale lengths are $a=1.49$ kpc, $b=0.58$ kpc, $c=0.40$ kpc.
The coordinates $x$ and $y$ span the galactic disk plane,
whereas  $z$ is perpendicular to it. The galactic disk is
described by a ``thin'' and a ``thick'' component (Gilmore, Wyse,
Kuijken \cite{gwk}). For the ``thin'' disk we adopt the following
density distribution
\begin{equation}
\rho_D(R,z) = \frac {\Sigma_0 } {2H} ~e^{-|z|/H}~e^{-(R-R_0)/h}~,
\label{rho_d}
\end{equation}
where the local projected mass density is $\Sigma_0 \simeq
25~M_{\odot}$ pc$^{-2}$, the scale parameters are $H\simeq 0.30$
kpc and $h\simeq 3.5$ kpc and $R_0$=8.5 kpc is the local
galactocentric distance. Here $R$ is the galactocentric distance
in the plane. For the ``thick'' component we consider the same
density law as in equation (\ref{rho_d})  with thickness $H\simeq
1$ kpc and local projected density $\Sigma_0 \simeq 50~M_{\odot}$
pc$^{-2}$. We also take into account the effect of the halo dark
matter component (in the form of MACHOs) which, as usual, is
assumed to follow the mass density profile
\begin{equation}
\rho_{\rm H }(r)=\rho_0^H\frac{a^2+R_0^2}{a^2+r^2}~,
\label{halo}
\end{equation}
where $\rho_0^H \simeq 7.9\times 10^{-3}~M_{\odot}$ pc$^{-3}$ is
the local dark matter density and $a\simeq 5.6$ kpc is the halo core
radius. We assume that halo lens objects have a mass $\simeq
0.4~M_{\odot}$ and constitute a fraction $f\simeq 0.5$ of the halo
dark matter as shown by microlensing observations (see e.g.
Alcock, Allsman, Alves et al. \cite{alcock}). \footnote{Very
recently, the MACHO group has reported conclusive results of
about $5.7$ years of microlensing observations directed towards the LMC. It
has been found that the MACHO halo fraction is now $f\simeq 0.2$
for a typical halo model; a fraction $f$ of up to one could still be 
compatible with observations (Alcock, Allsman, Alves et
al. \cite{macho2000}).}

As far as the mass distribution in globular clusters is
concerned, we assume a Plummer density profile given by
\begin{equation}
\rho_{\rm GC}(r)=\rho_0^{GC}\frac{1}{[1+(r/r_c)^2]^{3/2}}~,
\label{plummer}
\end{equation}
where $\rho_0^{GC}$ is the globular cluster central mass density
and $r_c$ the core radius.

Our galaxy should contain between a hundred million and a few
billions NSs; this follows from an extrapolation of the NS
birthrate (Narayan and Ostriker \cite{no}) and  from the number of
supernovae required to account for the heavy element abundances
in the Milky Way (Arnett, Schramm and Truran \cite{ast}). We will assume 
that the galactic bulge contains about
$10^9$ NSs. Furthermore, globular clusters are expected to contain many
NSs which are believed to be a different population with
respect to disk NSs. In the following, we will assume that each
globular cluster contain $10^3$ NSs (Grindlay and Bailyn
\cite{gb88}).

Rotating NSs are possible astrophysical sources of gravitational
waves in the frequency range of interferometric detectors such as
VIRGO and LIGO. The emission of gravitational waves, mainly at
the neutron star rotation frequency and at twice this frequency
(Zimmermann \cite{zim}), is a consequence of the 
asymmetric shape of the source caused by internal strong magnetic fields,
irregularities in the solid crust or to precessional motion.
It is easy to show that the gravitational wave amplitude from a
NS at distance $r$ from Earth, rotating with a period $P$, can be
expressed as
\begin{equation}
h_0\simeq4.2\times 10^{-24}\left(\frac{1{\rm ms}}{P}\right)^2
\left(\frac{1{\rm kpc}}{r}\right) \left(\frac{I}{10^{45}{\rm g~
cm^2}}\right) \left(\frac{\epsilon}{10^{-6}}\right), \label{h}
\end{equation}
where $\epsilon$ is the NS ellipticity and $I$ is the moment of
inertia. Since highly uncertain, the ellipticity $\epsilon$ is
the crucial parameter in this formula, and in order to
place limits on the gravitational wave maximum amplitude we shall
assume for $\epsilon$ the maximum value obtained by requiring
that most of the NS rotational energy is lost by gravitational
waves, i.e.
\begin{equation}
\epsilon_{\rm max}=\left(\frac{5c^5P^3\dot{P}}{512\pi^4
GI}\right)^{1/2}~, \label{ell}
\end{equation}
where $\dot{P}$ is the rate of deceleration and $I\simeq 10^{45}$ g
cm$^2$ is the typical momentum of inertia. For NSs in the
galactic bulge we obtain $\epsilon_{\rm max}\sim 10^{-4}$, since
for these objects typical values of $P$ and $\dot{P}$ are 0.1 s
and $10^{-15}$ s~s$^{-1}$, respectively. For globular cluster NSs
we obtain $\epsilon_{\rm max}\sim 10^{-8}$, taking for $P$ and
$\dot{P}$ the values 10 ms and $10^{-18}$ s~s$^{-1}$,
respectively (see e.g. Taylor, Manchester and Lyne \cite{tay} and
also Camilo, Lorimer, Freire et al. \cite{clf}).
\label{introduction}

\section{Microlensing of gravitational waves emitted by rotating NSs}

\subsection{NSs in the galactic bulge}

We first consider  NSs located in the galactic bulge microlensed
by the central black hole with mass $m\simeq 2.6\times
10^6~M_{\odot}$ (Genzel \cite{genzel98}, Ghez, Klein, Morris and
Becklin \cite{gkmb}). The corresponding Einstein radius is
between $R_E\simeq 6 - 20 \times 10^{16}$ cm, depending on
the distance of the source from the galactic center. Assuming that
the NS impact parameter is $d\le R_E$ (implying an amplification
$A\ge 1.34$), we can use the density distribution law given by 
equation (\ref{rho_c}) to calculate the rate $\Gamma$ of
BH microlensing events. In this case, equation (\ref{gamma})
results in $\Gamma\simeq 4\times 10^{-3}$ yr$^{-1}$.
\footnote{If we also take
into account the contribution of the NSs in the galactic
disk behind the central black hole, $\Gamma$ increases about
$10\%$.} From equation (\ref{duration}), the average time duration
of the NS gravitational wave amplification is $\Delta T\simeq
2\times 10^2$ yr, which is much longer than the typical
integration time for gravitational wave detection given in Table
\ref{table1} for the VIRGO detector. Such an extended event duration, in
addition to giving a substantially enhanced signal, should in
principle allow us to construct a Paczy\'{n}ski-like curve
(Paczy\'{n}ski \cite{pacz}) for microlensing of gravitational
waves. From equation (\ref{nev}), the number $N_{ev}$ of NSs
microlensed by the central black hole is therefore expected to be
$\simeq 1$ or higher, possibly up to a few tens, if one takes into
account the mass segregation effect and/or a $r^{-7/4}$ mass
density profile near the galactic center. \footnote{In
particular, taking a $r^{-7/4}$ profile up to $r\simeq 10^{-2}$
pc, which corresponds to the position of the last observed star
(Genzel \cite{genzel98}, Ghez, Klein, Morris and Becklin
\cite{gkmb}), we obtain a BH microlensing rate of $\Gamma\simeq 4\times
10^{-2}$ yr$^{-1}$.}
\begin{table*}
\caption{Lens Optical depth $\tau$, microlensing rate $\Gamma$ and
amplification time scale $\Delta T$ are shown for NSs located in
the galactic bulge (GB) and in three globular clusters (NGC7078,
NGC104 and NGC6380). The lens location BDH means that lenses are
bulge, disk and halo objects. The source location is given in the
third, fourth and fifth columns by using galactic coordinates $l$
and $b$ ($D$ is the source distance from Earth). For each source
location, results are given for microlensing by a central black
hole (first line) and for microlensing by stars and MACHOs (with
assumed mass $m\simeq 0.4~M_{\odot}$) distributed throughout the
galactic disk, bulge and halo (second line). The $\Gamma$ values
for microlensing by the central BH may be considered as lower
limits (see text).}
\medskip
\begin{tabular}{lllllllll}
\hline NSs Location& Lens Location&D (kpc)
&l$^{\circ}$&b$^{\circ}$ &Lens Mass ($M_{\odot}$)& $\tau$&
$\Gamma$
(yr$^{-1}$) & $\Delta T$  \\
\hline GB& Central BH& 8.5& 0&0 &$2.6\times 10^6$ & &$4.0\times 10^{-3}$&$2 \times 10^2$yrs \\
& ${\rm BDH}$ & 8.5 & & &0.4&$4.0\times 10^{-6}$ & $2.7\times 10^{-5}$& 60 days\\
 \hline
NGC7078& Central BH&9.7 & 65.0&-27.6 & $10^3$& &$3.7\times 10^{-7}$&8 yrs \\
                     & ${\rm BDH}$ &9.7& & &0.4 &$1.4\times 10^{-7}$ &$9.4\times 10^{-7}$ &60-260 days\\
 \hline
NGC104 & Central BH&4.6 & 305.89&-44.88 &$10^3$ & &$2.5\times 10^{-7}$&8 yrs \\
                     & ${\rm BDH}$ &4.6& & &0.4 &$3.9\times 10^{-8}$ &$3.5\times 10^{-7}$ &40-200 days\\
 \hline
 NGC6380 & Central BH& 4& 350.18&-3.414 & $10^3$& &$3.0\times 10^{-7}$& 3 yrs \\
                     & ${\rm BDH}$& 4& & &0.4 &$1.6\times 10^{-7}$ &$4.1\times 10^{-7}$ &40-120 days\\
\hline
\end{tabular}
\label{table2}
\end{table*}

Let us consider now BDH microlensing events by stars and MACHOs
with an average mass $m\simeq 0.4~M_{\odot}$ distributed throughout
the galactic disk, bulge  and halo, according to equations
(\ref{rho_c})-(\ref{halo}). \footnote{We note that towards the
galactic bulge, only a negligible number of microlensing
events are due to the halo component.} In this case, the duration
time scale of the gravitational wave amplification event is 
$\Delta T \simeq 60$ days, about the same value observed for
microlensing events of stars towards the galactic bulge (see e.g.
Alcock, Allsman, Alves et al. \cite{alcock}). Clearly, in this
case, even the next generation of gravitational wave detectors
will not allow the reconstruction of a Paczy\'{n}sky-like curve, but may
detect an enhanced gravitational wave amplitude with respect to the
non-microlensed NS. From equations (\ref{tau}) and (\ref{gamma})
we obtain $\tau\simeq 4.0\times 10^{-6}$ and $\Gamma \simeq
2.7\times 10^{-5}$ yr$^{-1}$, in agreement with the estimated
optical depth during the first year of optical microlensing
observations towards the galactic bulge. \footnote{In the first
year of the MACHO group observations, 45 events were detected by
observing $1.26\times 10^7$ stars towards the galactic bulge,
with an estimated optical depth $\tau_b=3.9^{+1.8}_{-1.2}\times
10^{-6}$ (Alcock, Allsman, Alves, et al. \cite{alcock_bulge}).}
We expect there are between $10^8$ and $10^9$ NSs in the galactic
bulge; therefore, if all these stars could be detected by 
gravitational wave experiments we would expect to see up to
$N_{ev}\simeq 10^5$ BDH microlensing events within 3 years of
integration time. Even if only a minor fraction of NSs could be
observed, one would expect to see a few gravitational wave
amplification events in a few years of observations.

From equation (\ref{h}) and from the discussion at the end of
Section \ref{introduction}, the gravitational wave amplitude for
bulge NSs is $h_0\simeq 3\times 10^{-26}$ and, taking into
account the microlensing effect, the maximum amplitude of the
metric perturbation on Earth is $h_{0}^ {\rm max}=\sqrt{A^{\rm
max}}~h_0$. It is evident that for impact parameters $d$ much
smaller than $R_E$ the amplification parameter $A^{\rm max}$
increases and, as occurs during the standard microlensing of
electromagnetic waves, the gravitational wave amplitude also increases
by a factor of $10$ or more, thereby facilitating its
detection.

\subsection{NSs in globular clusters}

Observations have shown that in our Galaxy at least 154 globular
clusters exist and that each of them contains about $10^3$ NSs
(Grindlay and Bailyn \cite{gb88}). Moreover, about $20\%$  of the
galactic globular cluster population (much more concentrated
towards the galactic center) are in a post-core-collapse (PCC)
phase and should contain a central massive black hole with mass
$\simeq 10^3~M_{\odot}$ (Chernoff and Djorgovski \cite{cd}).
Consequently, gravitational waves from NSs can be microlensed
either by the central black hole (BH events) or by lens objects
located along the line of sight and distributed through the
galaxy (BDH events), according to equations
(\ref{rho_c})-(\ref{halo}). In the following, we consider a few
PCC globular clusters.

First of all, let us consider the M15 (NGC7078) globular cluster
with a central black hole of mass $\simeq 10^3~M_{\odot}$
(Chernoff and Djorgovski \cite{cd}, Gebhardt and Fisher
\cite{gf}). We can evaluate the expected number of NSs behind the
Einstein circle of the central black hole by assuming they are
distributed like the stellar mass density, i.e. following a
Plummer density profile. By using equation (\ref{plummer}) with
$r_c\simeq 0.2$ pc and $\rho_0^{GC}\simeq 5.1\times 10^6$
$M_{\odot}$ pc$^{-3}$, so that the total mass enclosed within
about $10$ pc is $\simeq 4.8\times 10^5$ M$_{\odot}$
(Gebhardt and Fischer \cite{gf}), the obtained BH microlensing
rate is $\Gamma \simeq 3.7 \times 10^{-7}$ yr$^{-1}$.
\footnote{If we consider a King model for M15 with tidal radius
$r_t = 1.3$ pc, core radius $r_c = 0.1$ pc, central density
$\rho_0^{GC} = 4.2\times 10^6$ M$_{\odot}$ pc$^{-3}$ and of the
same total mass (Gebhardt and Fisher \cite{gf}, Lehmann and
Scholz \cite{ls}) instead of a Plummer model, $\Gamma$ does not
change substantially.} However, if we take into account the mass
segregation effect and/or a mass distribution law more
concentrated towards the center of the PCC globular cluster (with
an $r^{-7/4}$ profile), the lensing rate may be a factor of $10$
higher. In either case, the average amplification time scale is found
to be $\Delta T\simeq 8$ yr.

Let us now consider NSs in M15 microlensed by lenses in the Galaxy
displaced along the line of sight to M15. By using equations
($\ref{tau}$) and ($\ref{gamma}$), it is possible to calculate
the NS optical depth and the BDH microlensing rate, which turn out
to be $\tau\simeq 1.4\times 10^{-7}$ and $\Gamma\simeq 9.4\times
10^{-7}$ yr$^{-1}$, respectively. The amplification time scale of
the emitted gravitational waves is $\Delta T\simeq 60$
days for MACHOs belonging to the halo population and $\Delta
T\simeq 260$ days if the lenses are disk objects.

There are at least two other PCC globular clusters closer to
Earth than M15, and therefore more suitable for this kind of
observation: NGC104 (47 Tucane) and NGC6380, at distance of 4.6
kpc and 4 kpc, respectively. For the first globular cluster,
assuming a Plummer density profile with $r_c\simeq 0.50$ pc and
$\rho_0^{GC}\simeq 1.1\times 10^{5}$ $M_{\odot}$ pc$^{-3}$, and
considering gravitational waves from NSs microlensed by the
central black hole, we get a BH event rate $\Gamma\simeq 2.5 \times
10^{-7}$ yr $^{-1}$ and an amplification time scale $\Delta T
\simeq 8$ yr. For the microlensing events due to stars and/or
MACHOs distributed throughout the Galaxy, we get $\tau\simeq
3.9\times 10^{-8}$ and $\Gamma \simeq 3.5\times 10^{-7}$
yr$^{-1}$, while the average event duration is $\Delta T\simeq
40$ days for MACHOs in the galactic halo and $\Delta T\simeq 200$
days for lenses in the galactic disk.

Finally, we considered the PCC globular cluster NGC6380 described
with a Plummer model with $r_c\simeq 0.40$ pc and
$\rho_0^{GC}\simeq 4.4\times 10^{4}$ $M_{\odot}$ pc$^{-3}$. In the
case of BH microlensing events we get $\Gamma\simeq 3\times
10^{-7}$ yr$^{-1}$, corresponding to an average amplification time
scale $\Delta T\simeq 3$ yr. For BDH microlensing events we get in
this case $\tau\simeq 1.6\times 10^{-7}$ and $\Gamma \simeq
4.1\times 10^{-7}$ yr$^{-1}$, while the event duration is $\Delta
T\simeq 40-120$ days.
These results are summarized in Table \ref{table2}.

\section{Discussion and conclusions}

In this paper, we have discussed the astrophysical implications of
gravitational microlensing of gravitational waves emitted by
rotating NSs located in the galactic bulge and in a few globular
clusters. We consider this subject because with the
next generation of gravitational wave interferometric detectors,
it should be possible to obtain information on the galactic NS
population by studying emitted gravitational waves. As shown
in equation (\ref{h}), the amplitude of these waves can be
expressed in terms of the NS rotation period $P$, distance $r$ to
Earth, moment of inertia $I$ and ellipticity $\epsilon$ which we
estimate by equation (\ref{ell}). With the assumptions we made for
$P$, $\dot{P}$ and $\epsilon_{max}$, the amplitude of the
gravitational waves emitted by NSs located in the galactic bulge
and in globular clusters can be expressed, respectively, as
\begin{equation}
h_0^{\rm Bulge}\simeq
2.5\times10^{-25}\left(\frac{\rm{kpc}}{r}\right), ~~~h_{0}^{\rm
GC} \simeq7.9\times10^{-26}\left(\frac{\rm{kpc}}{r}\right)~.
\end{equation}
The detectability of gravitational waves from NSs by VIRGO-like
detectors was studied by many authors (see e.g. Shutz
\cite{schutz}, Gourgoulhon and Bonazzola \cite{gb}). These
authors pointed out that any pulsar (with $P\simeq 0.1$ s)
emitting gravitational waves with amplitude $h_0$ greater than
$10^{-26}$ in the frequency bandwidth where the sensitivity of
the detector is better than
$10^{-22}$ Hz$^{-1/2}$, can be detected in $3$ years of
integration time (see Table \ref{table1}). With this threshold,
it is possible to detect the gravitational
waves emitted by rotating NSs within about $25$ kpc, using the VIRGO 
detector. Clearly,
when we take into account the gravitational microlensing effect,
the amplitude of the gravitational waves is increased, thereby
facilitating their detection. As is shown by {\it
standard} microlensing observations towards the galactic bulge,
the amplification factor $A^{max}$ may be very high. In fact,
there are events with $A^{max}\simeq 70$ (Alcock, Allsman, Alves,
et al. \cite{alcock_bulge} and MACHO Project Home Page
\cite{web}), although the mean amplification averaged since 1993
to 1999 is $A^{max}\simeq 4$.
\begin{table}
\caption{The gravitational waves amplitude detectable by VIRGO in
$3$ yr of integration, given as a function of the frequency $f$
(see also Gourgoulhon and Bonazzola 1996).}
\begin{center}
\begin{tabular}{|c|c|c|}
\hline f (Hz)&Sensitivity (Hz$^{-1/2}$)& Detectable Amplitude\\
\hline 10&10$^{-21}$&10$^{-25}$\\
30&10$^{-22}$&10$^{-26}$\\
100& $ 3\times 10^{-23}$& $3\times10^{-27}$\\
1000& $ 3 \times 10^{-23}$ & $3\times10^{-27}$\\
\hline
\end{tabular}
\end{center}
\label{table1}
\end{table}

It is also easy to understand that the detection and the study of
the gravitational waves emitted from NSs can give remarkable
information regarding the population of pulsars inside the Galaxy and
globular clusters, providing a way to test current evolutionary
theories. Indeed, with the present astronomical observation
techniques it is not possible to distinguish objects behind the
massive central black holes and/or in the inner regions of the
galactic and globular cluster centers. However, 
gravitational wave detectors offer, in principle, a powerful tool
for this kind of astronomical investigation. If the gravitational
wave signal is higher than the background noise or if it is
sufficiently enhanced, for example by the microlensing effect, we
should be able to observe NSs which cannot be seen by any
other observational method. It is also true that the present
generation of interferometric detectors lacks sufficiently high
angular resolution to distinguish, as different gravitational
waves sources, NSs which are very close to each other. For
example, Jaranowki and Kr\`olak (\cite{jk}) showed that
VIRGO is expected to have an angular resolution $\Delta \Omega
\simeq 5.6\times 10^{-8}$ sr (corresponding to an angular
separation $\Delta \theta \simeq 27.5~\arcsec$), i.e. in the
center of our Galaxy, two NSs, with a separation distance $1$ pc,
should be observed as different sources. Otherwise, advanced LIGO
detectors should have a better angular resolution ($\Delta \Omega
\simeq 7.9\times 10^{-10}$ sr corresponding to $\Delta \theta
\simeq 3.3~\arcsec$), so that NSs $0.1$ pc apart could be resolved.
The next generation of interferometric
detectors, combined with a network analysis of the gravitational
wave signal, will likely increase the angular resolution which can be 
obtained. In general, two images are seen as two resolved sources if the angular
resolution of the telescope is below their angular separation,
$\Delta \theta _l$, whose expression is
\begin{equation}
\Delta \theta _l \simeq 2 \frac{R_E}{s}\sqrt{u^2+4}~.
\label{angular_separation}
\end{equation}
In the case of the black hole at the galactic center, the previous
equation results in $\Delta \theta _l\simeq 2.1\arcsec$, taking a
NS at a distance of $1$ kpc from the central black hole. Two
images separated by $2.1 \arcsec$ should be resolvable, for
example, by optical telescopes, however leaving open the problem
of detecting the lensing effect despite the large number of
background stars. In the case of gravitational
wave detectors, the lensed object is seen as a gravitational
microlensing event (i.e. gravitational wave amplification)
due to the more rough angular resolution. In the case of the
globular cluster NGC6380, the angular separation turns out to be
$\Delta \theta _l\simeq 9 \times 10^{-3} ~\arcsec$. Therefore, a
lensed source is seen as an amplitude amplification
(i.e. as a microlensing event).

We also note that recently Ruffa
(\cite{ruffa}) studied the gravitational lensing of
gravitational waves from a completely different point of view. 
He considered gravitational wave amplification as a
consequence of the Fraunhofer diffraction by a ring of radius
$R_E +\Delta R_E$ centered at the lens. In the case of gravitational waves 
from NSs
amplified by the black hole at the galactic center he found an
expected amplification $A=1.7\times 10^4$, which could significantly  aid 
observations. The problem with this analysis is the requirement for the 
perfect alignment of the lens and the source with the observer, which
should be an extremely unlikely situation in astrophysics. In a
following  paper (De Paolis, Ingrosso and Nucita \cite{dni}),
Ruffa's treatment is generalized to the situation of a non-aligned 
system in which the lens pass at some impact parameter
with respect to the line of sight to the source.

We emphasize that detecting such kind of microlensing events, in
addition to giving us a new test of the General Theory of Relativity,
should allow us to increase our knowledge of the galactic center
region and to constrain the NS distribution in the galactic bulge,
since the only possibility of detecting NSs behind the central black
hole is through the emitted gravitational waves. Such  gravitational 
events provide information about the nature
of the lensing object, allowing us, in principle, to discriminate between a
massive black hole and, for example, a compact star cluster (Maoz
\cite{maoz}) or a neutrino star (Munyaneza, Tsiklauri and
 Viollier \cite{mtv}, Capozziello and Iovane
\cite{capozziello}) at the galactic center.

Detecting microlensing events of NSs in globular clusters is also
important in order to obtain information about the evolution of a
globular cluster, i.e. about the PCC phase and/or the mass segregation
effect.

\end{document}